\documentclass[12pt]{article}
\usepackage{amsmath,amssymb,amsthm,amsxtra,overpic,bbm,bm,epsfig,ulem,multirow}
\usepackage{color,cite}
\usepackage{epstopdf}
\usepackage{booktabs}
\usepackage{footnote}
\usepackage{rotating}
\usepackage{hyperref}
\usepackage{url}
\usepackage{cite}
\usepackage{tikz}
\usetikzlibrary{arrows,shapes,automata,backgrounds,petri}
\usetikzlibrary{decorations.pathreplacing,decorations.markings}
\usetikzlibrary{decorations.pathmorphing}
\usetikzlibrary{decorations.markings}
\usepackage{xcolor}
\usepackage{lineno}
\usepackage{graphicx}
\usepackage{subcaption}
\usepackage{diagbox}

\def\thefootnote{\fnsymbol{footnote}}

\usepackage{array}

\vspace{0.2cm}

\begin{document}

\begin{center}

{\Large\bf Comprehensive study of cosmogenic neutron production in large liquid scintillator detectors} 

\end{center}

\vspace{0.2cm}

\begin{center}
{\bf Yijian Jiang$^a$, Jie Cheng$^a$~\footnote{Email: chengjie@ncepu.edu.cn}, Haoqi Lu$^b$~\footnote{Email: luhq@ihep.ac.cn}, and Yaoguang Wang$^c$} 
\\
\vspace{0.2cm}
{$^a$School of Nuclear Science and Engineering, North China Electric Power University, Beijing 102206, China}\\
{$^b$Institute of High Energy Physics, Chinese Academy of Sciences, Beijing 100049, China}\\
{$^c$Shandong University, Jinan 250100, China}
\end{center}

\vspace{1.5cm}

\begin{abstract}

Neutrons produced by cosmic ray muons constitute a significant background for underground experiments investigating neutron oscillations, neutrinoless double beta decay, dark matter, and other rare event signals. This work benchmarks measured neutron yields and neutron multiplicities—with a focus on data from the Daya Bay Reactor Neutrino Experiment—against comprehensive simulations using three GEANT4 hadronic physics lists. These simulations are further refined via a TALYS-based adjustment of hadronic cross sections. For the BERT-based models, the adjustment reduces the discrepancy in the total neutron yield from about 20\% to approximately 6\%, while for the BIC-based models it improves the agreement from roughly 13\% to the sub-percent level ($\sim$0.3\%), indicating a markedly better consistency of the BIC-based models with the experimental data. Nevertheless, a clear tension persists: simulations systematically underproduce single-neutron events while overproducing multi-neutron events. The study establishes a reproducible benchmark for cosmogenic neutron modeling and proposes a targeted refinement strategy—including channel-specific reweighting and intranuclear cascade parameter tuning—to guide future model development.

\end{abstract}

\def\thefootnote{\arabic{footnote}}
\setcounter{footnote}{0}

\newpage


\section{Introduction}
\label{sec:intro}

Neutrons produced by cosmic ray muons pose a significant background for large liquid scintillator (LS) detectors in underground experiments, including searches for neutrino oscillations, neutrinoless double beta decay, dark matter, and other rare events. Precise neutron modeling is therefore essential for background suppression and signal identification. Earlier studies, such as that by Wang {\it et al.}~\cite{Wang:2001fq}, have provided the FLUKA-based predictions for the dependence of the neutron yield on muon energy, with related work also performed by Kudryavtsey {\it et al.}~\cite{Kudryavtsev:2003aua}.
Experimental measurements of the neutron yields have been reported by various experiments at different overburden depths. Initial studies at CERN~\cite{Hagner:2000} provide fundamental data on neutron production cross-sections. Underground experiments including KamLAND~\cite{KamLAND:2010spallation}, Borexino~\cite{Borexino:2013cosmo}, and Daya Bay~\cite{DayaBay:2017prd} have further characterized neutron yields in their respective detector configurations. Additional measurements have been conducted by other groups~\cite{aglietta1989neutron,hertenberger1995muon,Boehm2000,Araujo:2008Boulby,Blyth2016,Reichhart2013}. Beyond total yields, neutron multiplicity distributions--which probe the underlying hadronic dynamics--have been studied at Boulby~\cite{Araujo:2008Boulby}, Borexino~\cite{Borexino:2013CosmoBG,Meindl:2013Thesis}, Daya Bay~\cite{DayaBay:2017prd}, and in a different interaction channel by T2K~\cite{T2K:2025NCnMult}.

Despite these experimental efforts, a systematic comparison between measured neutron observables and detailed simulations remains limited.
As an example, the Daya Bay~\cite{DayaBay:2017prd} have performed a precise measurement of the neutron production rate in LS at three different values of average muon energy with relative systematic uncertainties below 10\%. Corresponding predictions are calculated using both FLUKA~\cite{FLUKA:web} and GEANT4~\cite{GEANT4:web} simulation tools. However, the simulation results exhibit a discrepancy of $\sim$ 20\% when compared to the measured values. In other experiments, such as KamLAND~\cite{KamLAND:2010spallation}, the discrepancy between simulation and measurement also reaches around 30\% or more.
Discrepancies persist between default simulation frameworks and experimental data, especially in modeling hadronic interactions relevant for cosmic ray muon interactions. This gap impedes the refinement of simulation tools and constrains background predictions for future experiments, such as the Jiangmen Underground Neutrino Observatory (JUNO)~\cite{JUNO:2021vlw}. 

This work addresses the need for improved simulation accuracy by conducting a comprehensive benchmark study of cosmogenic neutron production in LS detectors, using the Daya Bay experiment as the primary reference.
We firstly employ a GEANT4-based simulation approach with three different hadronic physics lists to investigate the impact of hadronic interaction models on neutron production. Furthermore,  the simulations are enhanced by a TALYS~\cite{Koning:2023ixl}-based Monte Carlo adjustment scheme, which rescales GEANT4's inelastic cross sections for incident neutrons, protons, and gamma rays below 200 MeV. 
This methodology is motivated by the earlier study~\cite{Zugec2016} highlighting limitations in the GEANT4 hadronic models. In particular, the measurements of the reactions between neutron and $^{12}$C at CERN have shown that standard GEANT4 physics lists systematically underpredict neutron-$^{12}$C reactions. In contrast, the TALYS package with optimized parameters successfully reproduces the measured integral cross section for the $^{12}\mathrm{C}(n,p)^{12}\mathrm{B}$ reaction up to 10~GeV. These findings suggest that TALYS-based adjustments can substantially improve the accuracy of neutron production simulations. 
A comprehensive simulation study has been performed, covering the characterization of hadronic interaction models, and the estimations of neutron yields and multiplicity distributions. The improvements resulting from the TALYS-based cross section adjustments relative to the GEANT4 physics lists are quantitatively evaluated. Finally, the simulation results are compared to the Daya Bay measurements~\cite{DayaBay:2017prd} to calibrate the hardonic interaction models. 
Through comparisons between measurements and simulations, this work not only establishes a reproducible benchmark for cosmogenic neutron modeling, but also identifies persistent discrepancies in neutron multiplicity distributions that guide future model development.


The remainder of this paper is structured as follows. 
Section~\ref{sec:Calcalation} details the simulation strategy for cosmic muons, covering the cosmic-ray muon flux, the \textsc{GEANT4}-based detector simulation, and the implementation of the \textsc{TALYS}-based hadronic interaction model.
Section~\ref{sec:production} focuses on the characteristics of cosmogenic neutron production, analyzing relevant inelastic cross sections and the neutron production mechanisms.
Section~\ref{sec:res} presents the results, comparing simulated and experimental neutron yields and multiplicity distributions.
Finally, Sect.~\ref{sec:sum} summarizes the main findings and provides conclusions.

\section{Strategy for simulations}\label{sec:Calcalation}

\subsection{Cosmic ray muon flux}

To compare simulations with experimental data, this work employs measurements from the Daya Bay experiment~\cite{DayaBay:2017prd}, specifically using its LS detectors positioned in an experimental hall (EH1). This location is both close to the reactor core and situated at a vertical overburden of 250 meters-water-equivalent (m.w.e.).
In our simulations, the cosmic ray muon flux has been taken from the Ref.~\cite{DayaBay:2017prd} to ensure the muon spectrum accurately represents the experimental conditions. The following is a brief summary of the flux calculation based on Refs.~\cite{DayaBay:2017prd,Guan:2015vja}.

\begin{figure}[!tb]
    \centering
    \includegraphics[width=0.9\linewidth]{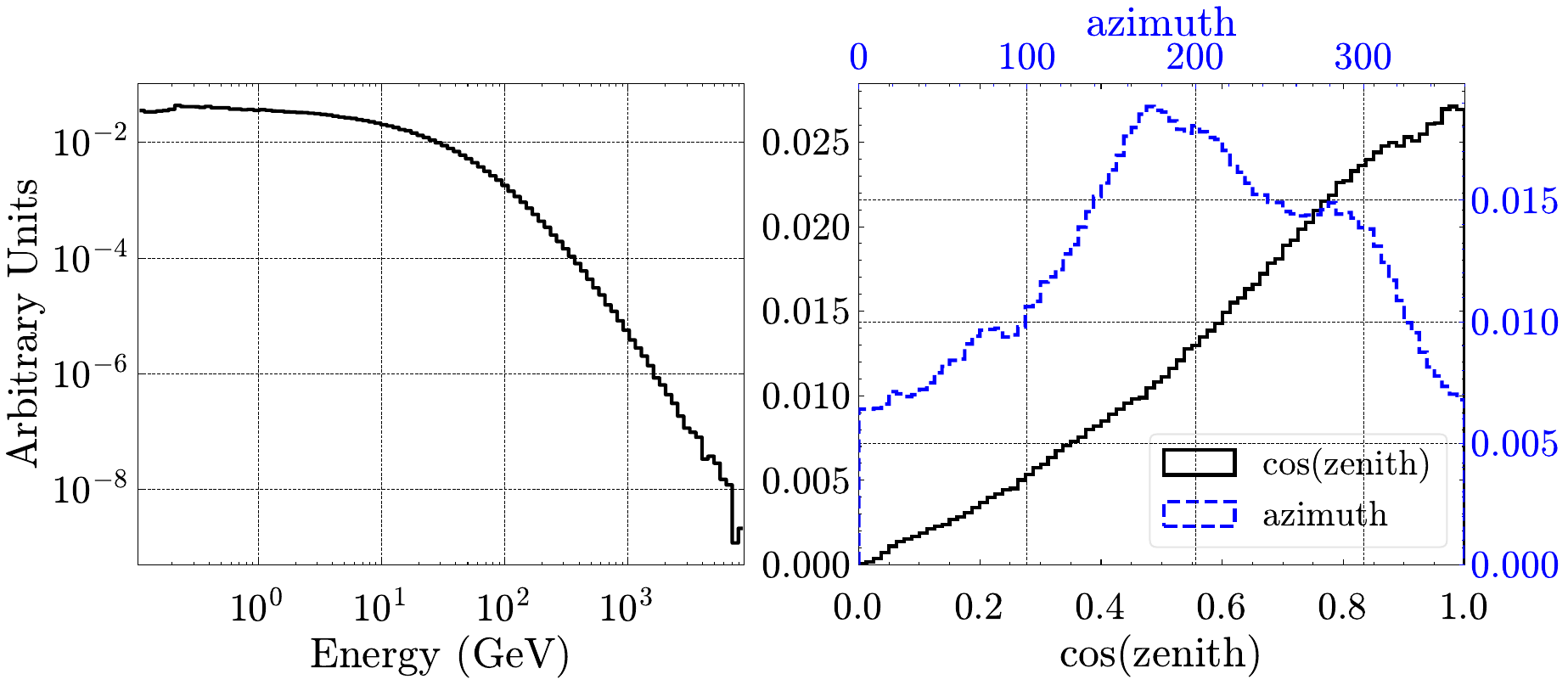}
    \caption{Energy spectrum and the angular distributions of the simulated muons. Left panel: normalized energy spectrum of muons, shown as a probability density with a logarithmic y-axis. Right panel: angular distributions of the muon directions. The black curve shows the probability density as a function of $\cos(zenith)$ (bottom x-axis), while the blue curve shows the azimuth angle distribution (top x-axis). Both angular distributions are normalized to unit area.}
    \label{fig:muon_init}
\end{figure}

The muon flux at sea level is well-modeled by Gaisser’s formula~\cite{Gaisser:CRPP,PDG:2004}. For the Daya Bay site, this formula has been modified to better describe the spectrum at low energies and large zenith angles~\cite{Guan:2015vja}, yielding fluxes consistent with previous cosmic ray muon measurements~\cite{Jokisch:1979gh, Kremer:1999sg, L3:2004sed, Rastin:1984nu, Ayre:1975qi}. To model site-specific shielding, a digitized mountain profile is generated from local topographic maps. The MUSIC code~\cite{Antonioli:1997, Kudryavtsev:2009} is then used to propagate muons from the top of this profile, calculating the muon path length through rock. The MUSIC simulation results give a muon rate of 1.27 Hz/m$^{2}$ and an average muon energy of 63.9 GeV.
Figure~\ref{fig:muon_init} shows the muon energy and angular distributions at the EH1. Here, the zenith angle is defined as a muon's angle from the vertical, and the azimuth angle is the compass bearing from true North. 

\subsection{GEANT4-based detector simulation}

\textsc{GEANT4} is a general-purpose Monte Carlo simulation toolkit developed by CERN, widely used in high-energy and nuclear physics for modeling the passage of particles through matter. It provides a flexible framework to simulate a variety of physical processes, including electromagnetic, hadronic interactions and optical processes, over a broad range of energies. In our simulations, \textsc{GEANT4} (version 11.2.1) has been employed to model muon-induced neutron production in LS detectors. Unless otherwise noted, all GEANT4 results presented in this work are based on version 11.2.1. For brevity and clarity, the version number will not be repeated in subsequent mentions.

\begin{figure}[!tb]
    \centering
    \includegraphics[width=0.5\linewidth]{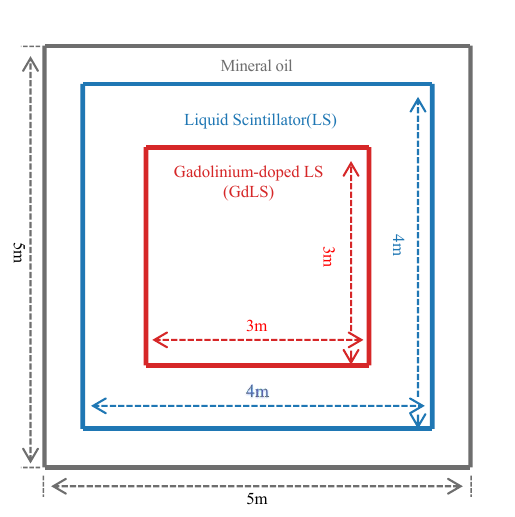}
    \caption{Schematic illustration of the detector geometry used in the simulation, showing the GdLS, LS, and mineral oil regions. The geometry of the water pool surrounding the LS detector is also modeled, though it is not illustrated here.}
    \label{fig:muon_geo}
\end{figure}

A detector geometry is constructed in \textsc{GEANT4} based on the Daya Bay experimental setup~\cite{DayaBay:2012fng}. 
The detector geometry, illustrated in Fig.~\ref{fig:muon_geo}, comprises three concentric cylindrical regions delineated by transparent acrylic vessels. The central target region (3 m in height and diameter) holds 20 tons of gadolinium-loaded liquid scintillator (GdLS). The material density ($\rho=0.86$ g/cm$^{3}$) is adopted from the measured value for the Daya Bay GdLS~\cite{DayaBay:2015kir}. This is surrounded by an equal-mass layer of pure LS and an outer mineral oil buffer. The entire assembly is housed within a 5 m $\times$ 5 m stainless steel vessel. The simulation focuses primarily on the GdLS region, where the majority of muon-induced neutrons are both produced and captured. Note that our simulation also includes muon detectors, sever as an water Cherenkov detector, in addition to the primary LS detector setup. 

\begin{figure}[!tb]
    \centering
    \includegraphics[width=0.9\linewidth]{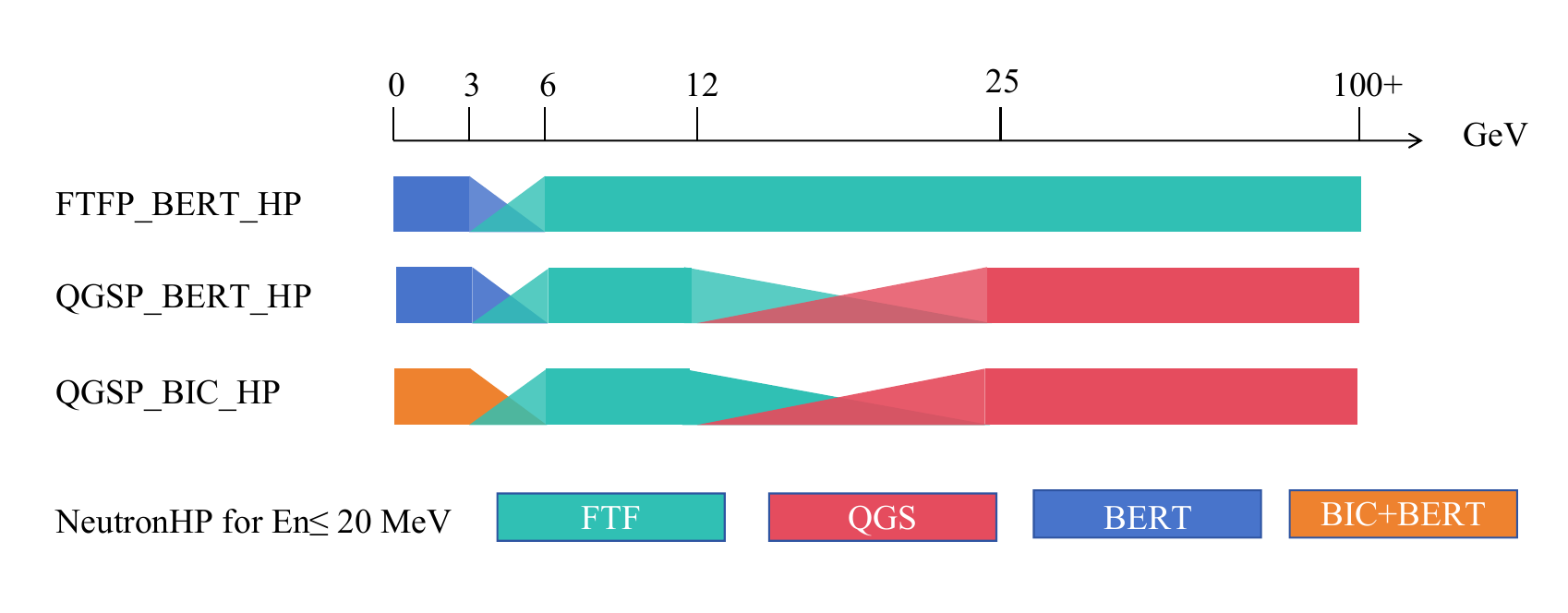}
    \caption{Model composition and energy ranges of the three \textsc{GEANT4} hadronic physics lists used in this work: \textbf{FTFP\_BERT\_HP}, \textbf{QGSP\_BERT\_HP}, and \textbf{QGSP\_BIC\_HP}. The shaded regions indicate smooth transitions between cascade and string models.}
    \label{fig:model_composition}
\end{figure}

To investigate the model dependence of hadronic interactions, our simulations employ three representative \textsc{GEANT4} hadronic physics lists: \texttt{FTFP\_BERT\_HP}, \texttt{QGSP\_BERT\_HP}, and \texttt{QGSP\_BIC\_HP}. 
Figure~\ref{fig:model_composition} illustrates the energy-dependent composition of these hadronic physics lists, based on the reference definitions provided in the \textsc{GEANT4} Physics List Guide~\cite{GEANT4:PLGuide}. Some comments on the Fig.~\ref{fig:model_composition} are helpful.
\begin{itemize}
    \item In the low- to intermediate-energy regime ($E < 6$~GeV), the intranuclear cascade is simulated using either the Bertini (BERT) or Binary (BIC) cascade model. As indicated by the transition regions in Fig.~\ref{fig:model_composition}, the simulation employs a linear transition in probability between the cascade models and the Fritiof (FTF) string model (typically between 3 and 6~GeV) to ensure continuity in physical observable.
    \item At higher energies, hadronic interactions are modeled using string excitation and fragmentation frameworks. A key distinction exists between the \texttt{FTFP} and \texttt{QGSP} variants. In the \texttt{FTFP\_BERT\_HP} list, the FTF model is used exclusively to handle interactions from the cascade transition region up to the highest energies. Conversely, the \texttt{QGSP} configurations adopt a multi-stage approach: the FTF model serves as an intermediate bridge (typically covering 6-25~GeV) to fill the gap between the cascade regime and the Quark-Gluon String (QGS) model, which dominates at very high energies ($>25$~GeV).
    \item The \texttt{QGSP\_BIC\_HP} configuration utilizes a hybrid scheme in the cascade region: the BIC Cascade model describes the interactions of primary nucleons ($p$ and $n$), while the BERT model handles other particles, such as pion ($\pi$), kaon ($K$), and $\gamma$. This distinction allows for a more detailed treatment of nucleon-induced spallation processes in the intermediate-energy regime.
    \item All three selected lists incorporate the Neutron High-Precision (HP) package (denoted by the \texttt{\_HP} suffix). This module utilizes evaluated data-driven cross sections for neutrons below 20~MeV, providing a high-fidelity description of low-energy neutron transport, moderation, and capture processes, which are critical for reproducing thermal-neutron yields in LS detectors.
\end{itemize}

Note that these physics lists are widely adopted in neutrino and cosmic-ray studies because they provide complementary modeling of hadron-nucleus interactions across different energy ranges. In this work, they are specifically chosen because they span the dominant sources of hardonic interaction model uncertainties in \textsc{GEANT4}: \texttt{FTFP\_BERT\_HP} and \texttt{QGSP\_BERT\_HP} differ in their high-energy string models (FTF vs.\ QGS), while \texttt{QGSP\_BIC\_HP} replaces the BERT cascade with the BIC Cascade at intermediate energies. Comparing these three lists therefore allows a systematic assessment of both string-model and intranuclear-cascade modeling uncertainties that directly impact the neutron yields.

\subsection{TALYS-based hardonic interaction model}
\label{sec:talys}
The \textsc{TALYS}  package~\cite{Koning:2012} models nuclear reactions for projectiles—including $p$, $n$, $d$, $t$, $^3$He, $\alpha$, and $\gamma$—over energies from 1 keV to 200 MeV. It calculates cross sections, residual nucleus production, and emission spectra using optical, pre-equilibrium, and Hauser–Feshbach statistical models. 

This study employs \textsc{TALYS} (1.8) to provide hadronic interaction cross sections for $p-^{12}$C, $n-^{12}$C, and $\gamma-^{12}$C, complementing the \textsc{GEANT4} simulation. Because \textsc{TALYS} does not simulate $\pi^{\pm}-^{12}$C or $\mu-^{12}$C interactions, these processes are treated exclusively with \textsc{GEANT4} in this work. Version 1.8 is used to maintain consistency with the \textsc{TALYS}  version integral to the TENDL-2017 nuclear data library~\cite{TENDL2017}. TENDL-2017 supplies ENDF-formatted data from default and adjusted \textsc{TALYS} calculations, augmented by other sources. The library and its toolchain have undergone extensive verification, validation, and benchmarking against differential and integral experimental datasets in basic and applied nuclear physics. Another reason for this version choice is that while newer \textsc{TALYS}  releases include improvements—notably in photonuclear reactions and fission yields—the optical model parameters and pre-equilibrium descriptions for nucleon-induced inelastic scattering on light nuclei like \(^{12}\text{C}\) (at energies \(E < 200~\text{MeV}\)) are already well-established in version 1.8. We therefore use \textsc{TALYS} (1.8) as a robust 
physics baseline to evaluate the inclusive inelastic cross-section discrepancies observed in \textsc{GEANT4}. Note that the \textsc{TALYS} (1.8) is used throughout this work. Subsequent references to \textsc{TALYS} will omit the version number for clarity.

To incorporate the \textsc{TALYS}-based hadronic cross sections into the \textsc{GEANT4} simulations, we have noted that a single $\mu$-induced cascade comprises multiple individual $n$-producing vertices. Consequently, we implement a vertex-level reweighting scheme for each $\mu$ event. In this \textsc{TALYS}-based MC framework, the total neutron number for a $\mu$ event, denoted $N_{n}^{\mathrm{T}}$, is calculated as per  Eq.~\ref{eq:NnT} by reweighting each new $n$ generated vertex $i$ according to the parent particle type $j$ (for $j=n,p,\gamma$) and its kinetic energy $E_i$.

\begin{equation}
    N_{n}^{\mathrm{T}} = \sum_{i,j} N^{\mathrm{G}}_{i,j} \times W_{i,j}(E_i, j)
    \label{eq:NnT}
\end{equation}

where $N^{\mathrm{G}}_{i,j}$ denotes the number of neutrons produced at vertex $i$ by a parent particle of type $j$ (for $j=n,p,\gamma$) in the original \textsc{GEANT4} simulation. The corresponding weighting factor $W_{i,j}$ is derived from the exclusive new-$n$-generation inelastic cross sections ($\sigma_{in}$) and is defined as:

\begin{equation}
    W_{i,j} = 
    \begin{cases} 
      \dfrac{\sigma^{\text{T}}_{in, j}(E_i)}{\sigma^{\text{G}}_{in,j}(E_i)} & \text{for } j =n, p, \gamma\text{ on } ^{12}\text{C} \\
      1 & \text{otherwise}
   \end{cases}
   \label{eq:weight_factor}
\end{equation}

Where, $\sigma^{\text{G}}_{in}$ and $\sigma^{\text{T}}_{in}$ denote the \textsc{GEANT4} and \textsc{TALYS} exclusive inelastic cross sections for new $n$ generated reactions, respectively. Note that for vertices with $E_i > 200$ MeV (beyond \textsc{TALYS}'s range), we set
$W_{i,j} = 1$, using the original \textsc{GEANT4} result. What's more, if an exclusive new-$n$-generation channel predicted by the original \textsc{GEANT4} simulation is absent from \textsc{TALYS}, its corresponding $W_{i,j}$ is set to 1, leaving its cross section unchanged.  

This reweighting scheme adjusts the interaction rate for exclusive channels (governed by their exclusive $\sigma_{in}$) while preserving the final-state multiplicity and kinematics predicted by \textsc{GEANT4}.
Crucially, it does not substitute \textsc{GEANT4}'s exclusive channels—which rely on its internal de-excitation models (e.g., Fermi break-up)—with \textsc{TALYS} predictions event by event. Such a substitution would violate the kinematic consistency of the simulated cascade, as secondary particle momenta must be balanced with the specific nuclear remnant. Instead, our hybrid method renormalizes the reaction probability using more accurate nuclear data from \textsc{TALYS}.

\section{Characteristics of cosmogenic neutron production}
\label{sec:production}

In our \textsc{GEANT4}-based detector simulations, cosmic muons have been sampled above the detector geometry (Fig.~\ref{fig:muon_geo}), following the energy and angular distributions shown in Fig.~\ref{fig:muon_init}. The propagation of these muons through the full detector geometry has simulated using three representative GEANT4 hadronic physics lists: \texttt{FTFP\_BERT\_HP}, \texttt{QGSP\_BERT\_HP}, and \texttt{QGSP\_BIC\_HP}, as detailed in Sect.~\ref{sec:Calcalation}. The neutron yield is defined as the number of neutrons produced per incident muon, per unit path length, and per unit material density. To enable a direct comparison with Daya Bay measurements~\cite{DayaBay:2017prd}—specifically those involving GdLS—our simulation focuses on muons traversing the GdLS region. This corresponds to the experimental procedure where neutrons captured on Gd, following an identified muon in the LS, are selected. Consequently, we have analyzed the corresponding muon tracks through GdLS and the subsequent neutron production within this volume. 

Simulation results show that $\sim$61\% of muons depositing at least 20 MeV in the LS also traverse the GdLS. This fraction is consistent across the three physics lists and agrees with Daya Bay simulations~\cite{DayaBay:2017prd}. Furthermore, the average muon path length in the GdLS ($L_{\rm{avg}}$), a geometry-dependent parameter sensitive to the muon angular distribution, is about 206.3 cm for EH1 in our simulations. This value is also consistent across physics lists and compares well with the 204.1 cm obtained from Daya Bay simulations for EH1~\cite{DayaBay:2017prd}. Finally, we have verified the distribution of energy deposited in the LS by muons with a nonzero track length in the LS against Daya Bay measurement data~\cite{DayaBay:2017prd}; the spectral shapes are consistent between our simulations and the experimental data.
 
The following part of this section analyzes the characteristics of the resulting neutron production processes.
First, we compare inelastic reaction cross sections for neutron production in the GdLS region. Then, this analysis progresses from energy-dependent inelastic cross section predictions to an examination of dominant neutron multiplying reaction channels. Finally, exclusive cross sections are evaluated to quantify the model-dependent differences that directly impact for spallation neutron modeling.

\subsection{Inelastic cross sections}
The dominant contributors to $\mu$-induced neutron production in GdLS region are secondary $n$, $p$, $\gamma$, $\pi^{+}$, and $\pi^{-}$. These particles, generated primarily along muon tracks through electromagnetic and hadronic cascades, initiate the subsequent interactions with carbon nuclei that yield spallation neutrons. Therefore, a detailed characterization of the inelastic reaction cross sections ($\sigma_{in}$), for these projectiles interacting with $^{12}$C is essential for identifying the microscopic origins of spallation neutron and benchmarking model-dependent effects.

\begin{figure}[!tb]
    \centering
    \includegraphics[width=0.9\linewidth]{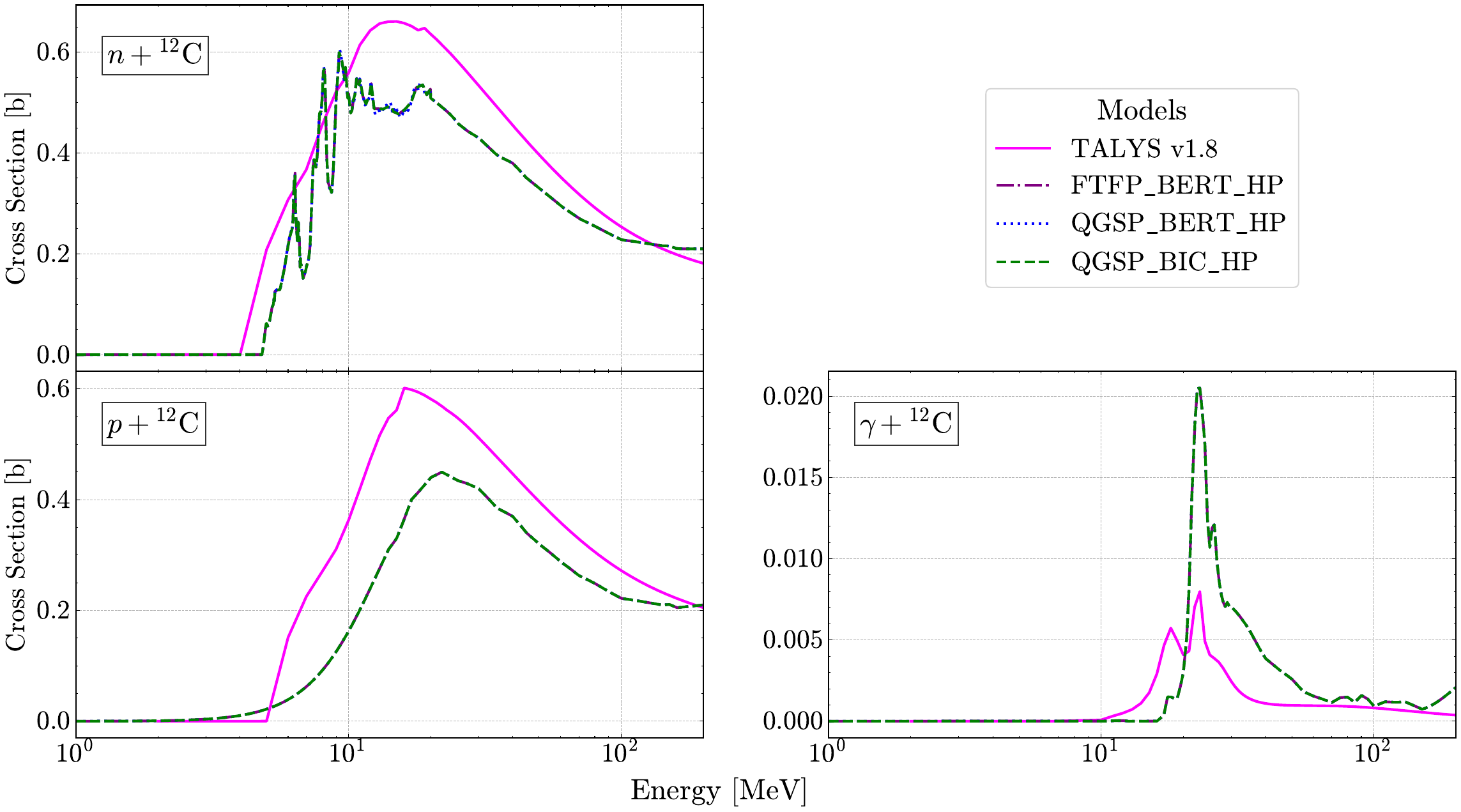}
    \caption{Comparison of inclusive inelastic cross sections ($\sigma_{in}(E)$) for $n$, $p$, and $\gamma$ interacting with $^{12}$C, obtained from \textsc{TALYS}  and \textsc{GEANT4}  hadronic models (\texttt{FTFP\_BERT\_HP}, \texttt{QGSP\_BERT\_HP}, and \texttt{QGSP\_BIC\_HP}). Note that charged pions are not included in this comparison as they are not supported by the \textsc{TALYS} framework.}
    \label{fig:g4_talys_inel_xs_compare}
\end{figure}

\begin{figure}[!tb]
    \centering
    \includegraphics[width=0.9\linewidth]{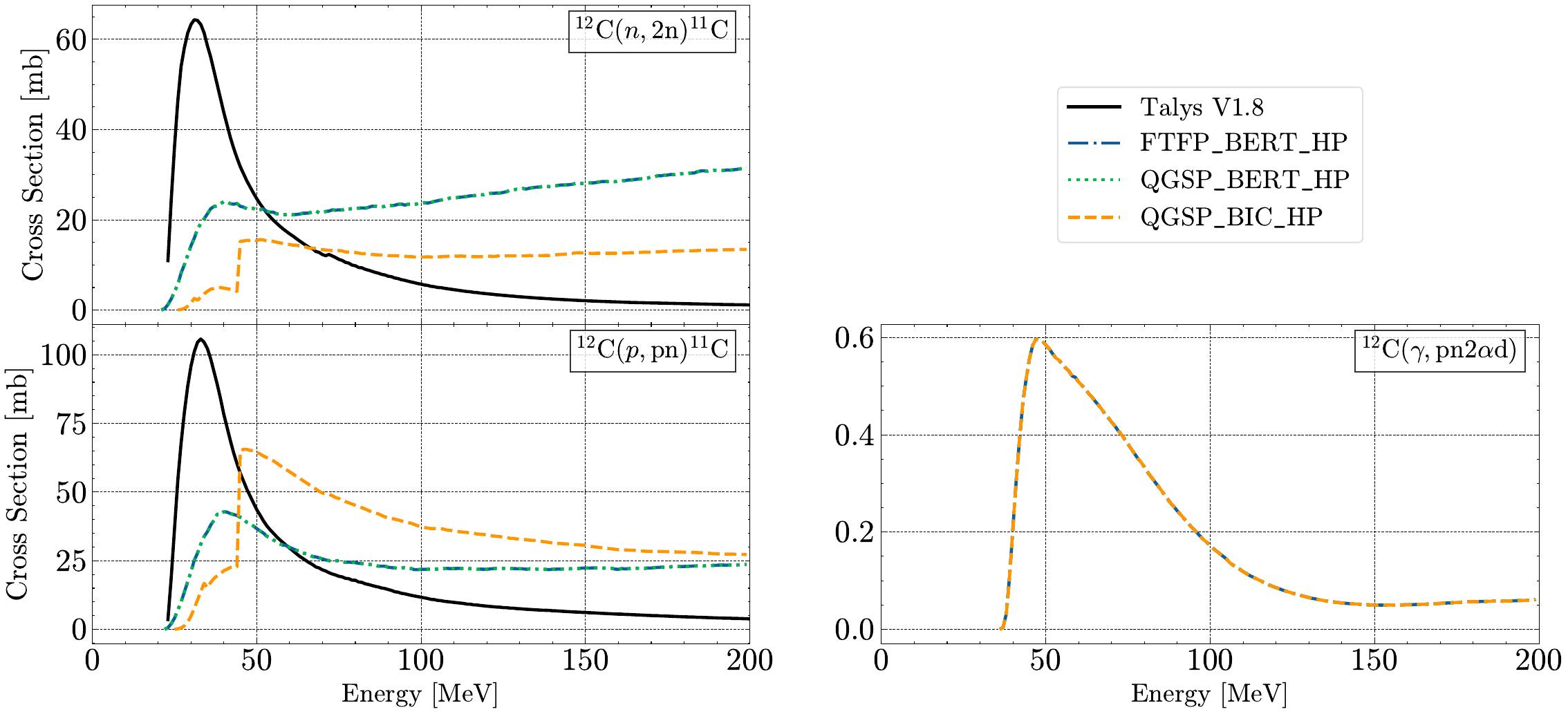}
    \caption{Comparison of the dominant neutron-multiplying inelastic reaction-channel cross sections for $n$, $p$, and $\gamma$ projectiles on $^{12}\mathrm{C}$ below 200 MeV. For each projectile, the reaction channel in which the total $n$ number in the final state exceeds that of the incident particle and has the highest occurrence probability in \textsc{GEANT4}  \texttt{FTFP\_BERT\_HP} is identified, and the corresponding cross sections are compared with those from \texttt{QGSP\_BERT\_HP}, \texttt{QGSP\_BIC\_HP}, and \textsc{TALYS}.}
    \label{fig:ratio_talys_geant4}
\end{figure}

The inclusive $\sigma_{in}$ is a fundamental quantity governing the neutron production rate. To investigate model dependencies, the $\sigma_{in}$ for the \textsc{TALYS}-supported projectiles, namely $n$, $p$, and $\gamma$, incident on $^{12}$C using \textsc{TALYS} , is compared to the predictions of three \textsc{GEANT4}  hadronic physics lists: \texttt{FTFP\_BERT\_HP}, \texttt{QGSP\_BERT\_HP}, and \texttt{QGSP\_BIC\_HP}. Because \textsc{TALYS} cannot model $\pi^\pm$ or $\mu^\pm$ as projectiles, the \textsc{TALYS}–\textsc{GEANT4} comparison is performed only for the $n$, $p$, and $\gamma$ channels, whereas the $\pi^\pm$ (and $\mu^\pm$) contributions are included solely in the full \textsc{GEANT4}–based evaluation of $\mu$-induced neutron production.
Since the incident energy simulated by \textsc{TALYS} is limited to 200 MeV,  Fig.~\ref{fig:g4_talys_inel_xs_compare} presents the comparison across this energy range. For the comparison of these \textsc{GEANT4} physical model lists, we have found that the inclusive $\sigma_{in}$ across the full energy range for these particles are nearly identical. This consistency can be understood as follows.
\begin{itemize}
    \item For $p$ and $\gamma$, the hadronic models BERT and BIC are used for $E\,<$ 200 MeV. Consequently, the cross sections from \texttt{FTFP\_BERT\_HP} and \texttt{QGSP\_BERT\_HP} are identical.
    \item For $n$, the HP model is applied below 20 MeV, while above this threshold the same BERT/BIC models used for $p$ and $\gamma$ are employed. Therefore, below 20 MeV the cross sections from all three lists coincide, whereas above 20 MeV the results from \texttt{FTFP\_BERT\_HP} and \texttt{QGSP\_BERT\_HP} are the same.
    \item The inclusive cross sections for $n$, $p$, and $\gamma$ from the BERT and BIC models are themselves very similar in this energy range. However, significant differences emerge in exclusive reaction channels, as will be shown in Fig.~\ref{fig:ratio_talys_geant4}.
\end{itemize}
Turning to the comparison between \textsc{TALYS} and \textsc{GEANT4}, the inclusive  $\sigma_{in}$ for $n$ and $p$ from \textsc{TALYS} and \textsc{GEANT4} exhibit  agreement in the spectral shape in Fig.~\ref{fig:g4_talys_inel_xs_compare}, but a systematic discrepancy in magnitude: \textsc{TALYS} values are $\sim$20\% ($n$) and 30\% ($p$) higher across this energy range. For $\gamma$, notable differences emerge: the \textsc{TALYS} cross section peak is shifted to lower energies and its overall magnitude is roughly 50\% lower than that of \textsc{GEANT4}. 

Given our focus on exclusive channels related to neutron generation, we have analyzed the exclusive cross sections of the dominant new-neutron-producing channels, where the final-state neutron count exceeds that of the incident particles.
A previous study~\cite{Cheng:2023zds} has calculated exclusive cross sections for $n$-induced reactions on $^{12}$C using \textsc{TALYS}, identifying $^{12}\mathrm{C}(n,2n)^{11}\mathrm{C}$ and $^{12}\mathrm{C}(n,2n\alpha)^{7}\mathrm{Be}$ as the dominant new-neutron-generation channels. Following this work, we have performed analogous \textsc{TALYS}  calculations for $p$- and $\gamma$-induced reactions on $^{12}$C. For $p-^{12}$C reactions, the dominant new-neutron-generation channels are $^{12}\mathrm{C}(p,n)^{12}\mathrm{B}$ and $^{12}\mathrm{C}(p,np)^{11}\mathrm{C}$. For $\gamma-^{12}$C reactions, the dominant channel is $^{12}\mathrm{C}(\gamma,2n2p)^{8}\mathrm{Be}$.

The exclusive cross sections for $n$-, $p$-, and $\gamma$-induced reactions on $^{12}$C are also obtained from \textsc{GEANT4}. Fig.~\ref{fig:ratio_talys_geant4} shows the energy dependence of these leading channels, selected from the highest-probability modes in the \textsc{GEANT4}  \texttt{FTFP\_BERT\_HP} physics list. Each panel, corresponding to a different projectile ($n$, $p$, or $\gamma$), compares the predictions of three \textsc{GEANT4}  physics lists with the available \textsc{TALYS}  calculations. For the $^{12}$C$(n,2n)^{11}$C and $^{12}$C$(p,pn)^{11}$C channels, the cross sections predicted by \texttt{FTFP\_BERT\_HP} and \texttt{QGSP\_BERT\_HP} are essentially identical across the 0$-$200~MeV. This indicates that, the BERT intra-nuclear cascade is dominant in this energy range. The choice of high-energy string model (FTF vs.\ QGS) has little impact on these particular channels. In contrast, \texttt{QGSP\_BIC\_HP}, which substitutes BIC for BERT at intermediate energies, yields significantly different cross sections, particularly in the tens-of-MeV region. This result demonstrates that within \textsc{GEANT4}, the treatment of the intra-nuclear cascade is more influential than the string model in determining the relative weights of specific $n$-producing channels. In the \texttt{QGSP\_BIC\_HP} physics list, the BERT model remains active for projectiles other than neutrons and protons. Therefore, the exclusive cross sections for $\gamma$-induced neutron production are the same across all three physics lists.

When compared to \textsc{TALYS}, systematic differences become apparent. For the $n$- and $p$-induced neutron examples, \textsc{TALYS}  predicts a pronounced maximum near threshold and generally larger inelastic strengths at low energies for the $^{12}$C$(n,2n)^{11}$C and $^{12}$C$(p,pn)^{11}$C channels than any of the three \textsc{GEANT4}  physics lists. 
For the $\gamma$-induced neutron case, \textsc{GEANT4} identifies a highly specific multi-fragment breakup channel, $^{12}$C$(\gamma,pn2\alpha d)$, as the most probable neutron-multiplying mode in \texttt{FTFP\_BERT\_HP}, and all three physics lists yield almost identical cross sections for this channel. \textsc{TALYS}, however, assigns negligible strength to this exclusive high-multiplicity final state within its Hauser--Feshbach framework. The presence of this channel in \textsc{GEANT4} and its practical absence in \textsc{TALYS} further highlights the differences in the deexcitation modeling of highly excited $^{12}$C between the two simulation tools.

Collectively, Figs.~\ref{fig:g4_talys_inel_xs_compare}–\ref{fig:ratio_talys_geant4} reveal that although \textsc{GEANT4} physics lists are self-consistent and reproduce the qualitative trends of dominant neutron-production channels, they underestimate low-energy inelastic strength relative to \textsc{TALYS}. The more detailed nuclear-structure and energy-dependence effects in \textsc{TALYS} motivate the hybrid rescaling scheme (Sec.~\ref{sec:talys}) developed to improve simulations of $\mu$-induced spallation neutrons.

\subsection{Production of cosmogenic neutrons}

Following the \textsc{GEANT4}-based simulation of cosmic muons, we have investigated the neutron-producing processes to quantify the impact of different hadronic models on secondary neutron generation. We isolate channels resulting in net neutron gain (final-state neutrons $>$ incident neutrons). For inelastic interactions on $^{12}$C, the fraction of events leading to new neutron neutron generation varies by projectile and \textsc{GEANT4} physics list. 

\begin{figure}[!tb]
    \centering
    \includegraphics[width=0.8\linewidth]{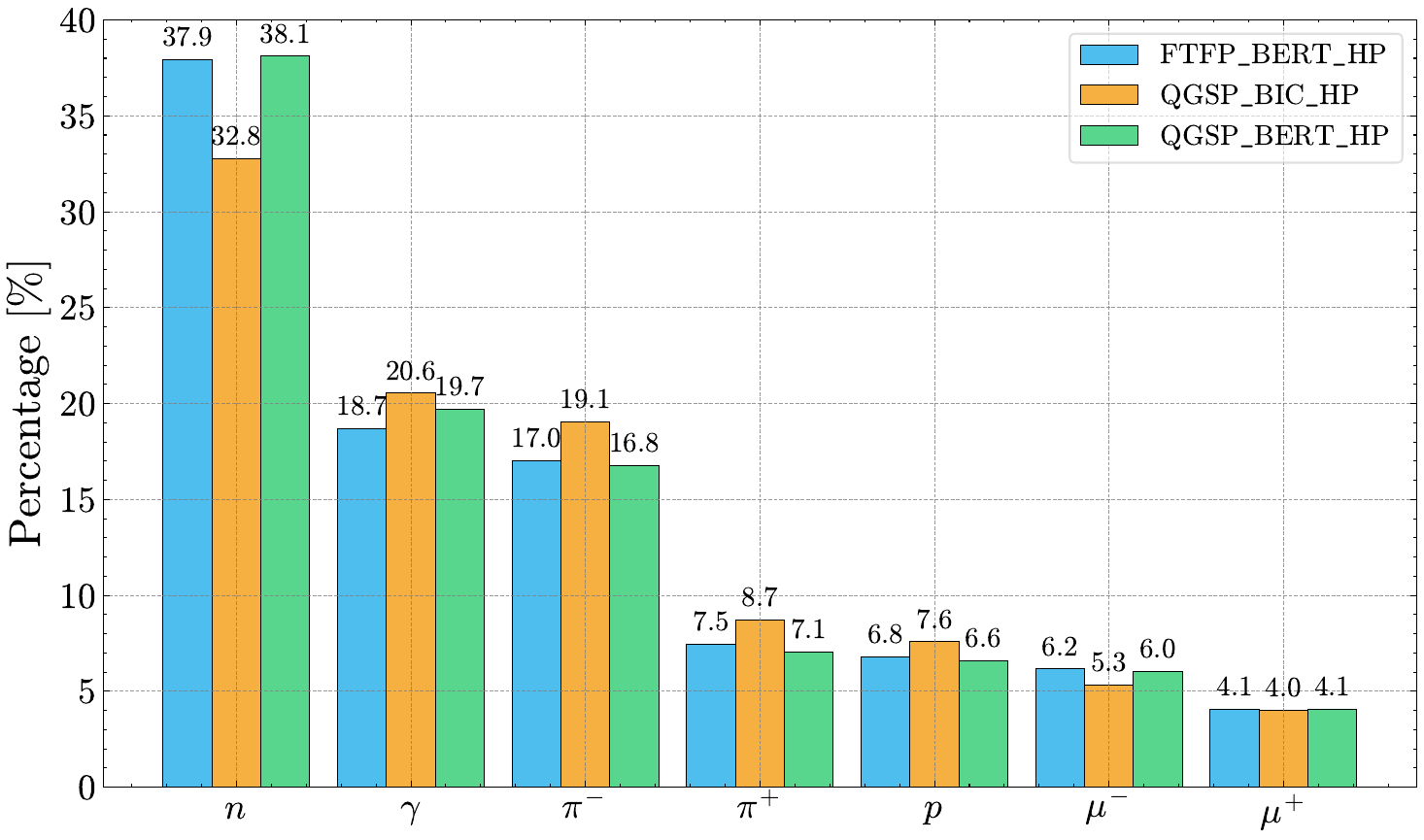}
    \caption{Distribution of parent particles responsible for neutron production under different hadronic physics lists. 
    The relative contributions (\%) of each parent type are compared among \texttt{FTFP\_BERT\_HP}, \texttt{QGSP\_BERT\_HP}, and \texttt{QGSP\_BIC\_HP}.}
    \label{fig:process_yield_ftfp}
\end{figure}

For the new-neutron-producing processes, Fig.~\ref{fig:process_yield_ftfp} presents the relative contributions of various parent particles to the new neutron production across the three \textsc{GEANT4} physics lists. The results reveal that $n$-, $\gamma$-, and $\pi^{-}$-induced reactions dominate the overall neutron yield, followed by processes induced by $\pi^{+}$, protons, and muons.
Comparison of the physics lists reveals a distinct behavior for \texttt{QGSP\_BIC\_HP}. While the two BERT-based models (\texttt{FTFP\_BERT\_HP} and \texttt{QGSP\_BERT\_HP}) exhibit nearly identical distributions, \texttt{QGSP\_BIC\_HP} predicts a \textbf{lower} relative contribution from $n$-induced reactions ($\sim$32.8\% vs. $\sim$37.9\%) and an \textbf{enhanced} contribution from charged pions and protons. This shift is consistent with the model definitions: \texttt{QGSP\_BIC\_HP} employs the BIC model for nucleons ($n, p$) but retains the BERT model for other particles. The difference suggests that the BIC treatment of nucleon-nucleus interactions alters the branching ratios of secondary cascades, effectively redistributing the production weight toward $\pi$- and $p$-mediated channels relative to the BERT model.

\begin{figure}[!tb]
    \centering
    \includegraphics[width=0.9\linewidth]{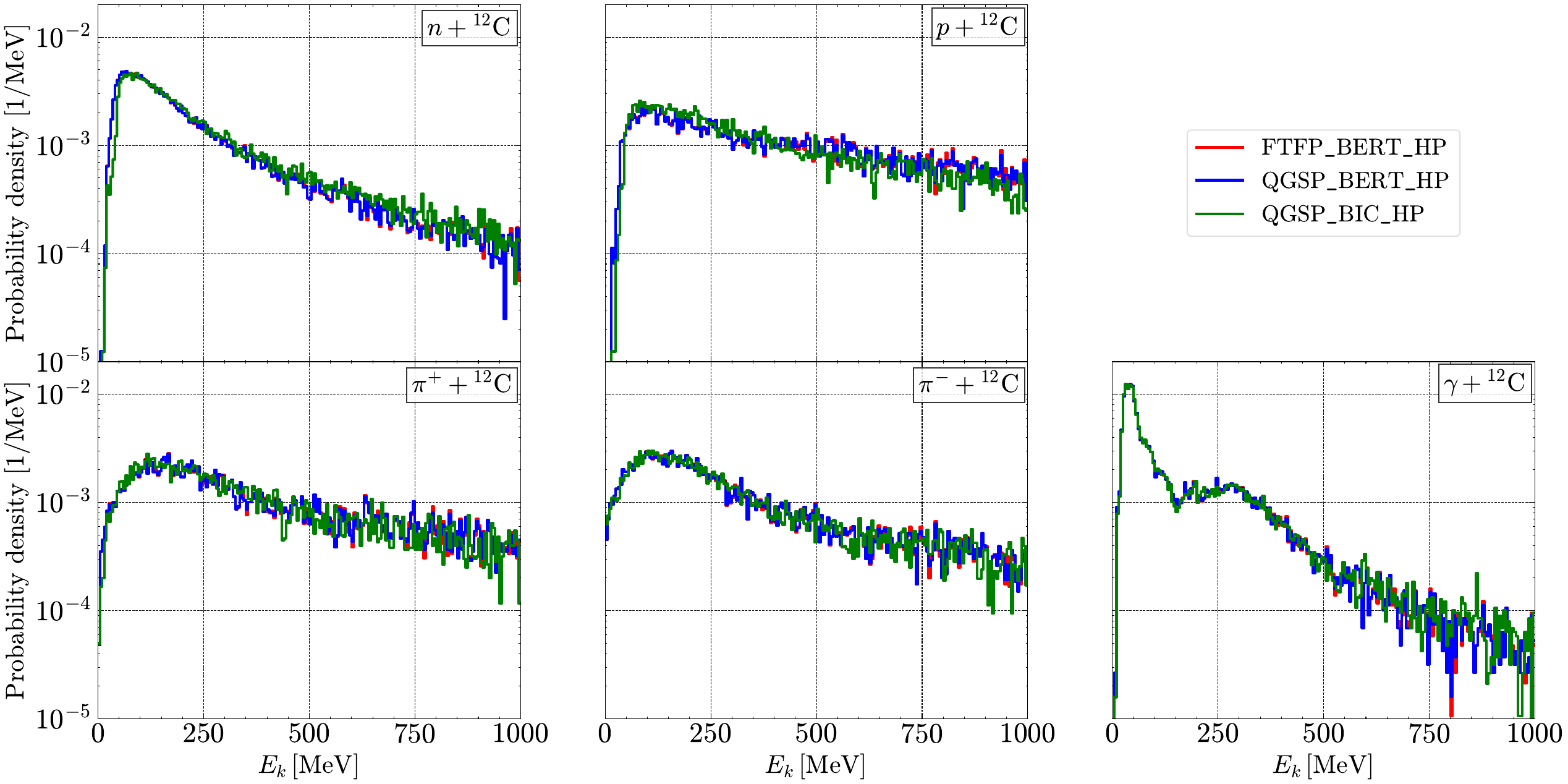}
    \caption{Parent-particle kinetic-energy spectra at the neutron-production vertex,
    resolved by initiating particle 
    ($n\!\to\!n$, $p\!\to\!n$, $\gamma\!\to\!n$ and $\pi^{\pm}\!\to\!n$)
    for three \textsc{GEANT4} physics lists.}
    \label{fig:parentKE_allPL}
\end{figure}

Next, we have examined the kinetic-energy distributions of the parent particles at the neutron-production vertex. 
Figure~\ref{fig:parentKE_allPL} compares the channel-resolved parent-energy spectra across the three physics lists. For all initiating particles ($n$, $p$, $\gamma$ and $\pi^\pm$), the spectra feature a prominent soft component below $\sim$150~MeV and a long high-energy tail extending up to the GeV scale. Hadronic parents $p$ (and, similarly, $\pi^\pm$) tend to exhibit somewhat harder high-energy tails than $\gamma$ and $n$, consistent with intranuclear-cascade kinematics and multi-nucleon emission in spallation reactions. The remarkable stability of these spectral shapes among the physics lists suggests that the variations in the relative contributions of parent particles for neutron productions (Fig.~\ref{fig:process_yield_ftfp}) are driven primarily by differences in \textbf{reaction probabilities (cross sections) and branching ratios}, rather than by large shifts in the accessible kinetic-energy phase space of the parent particles.

 \begin{figure}[!tb]
    \centering
    \includegraphics[width=0.8\linewidth]{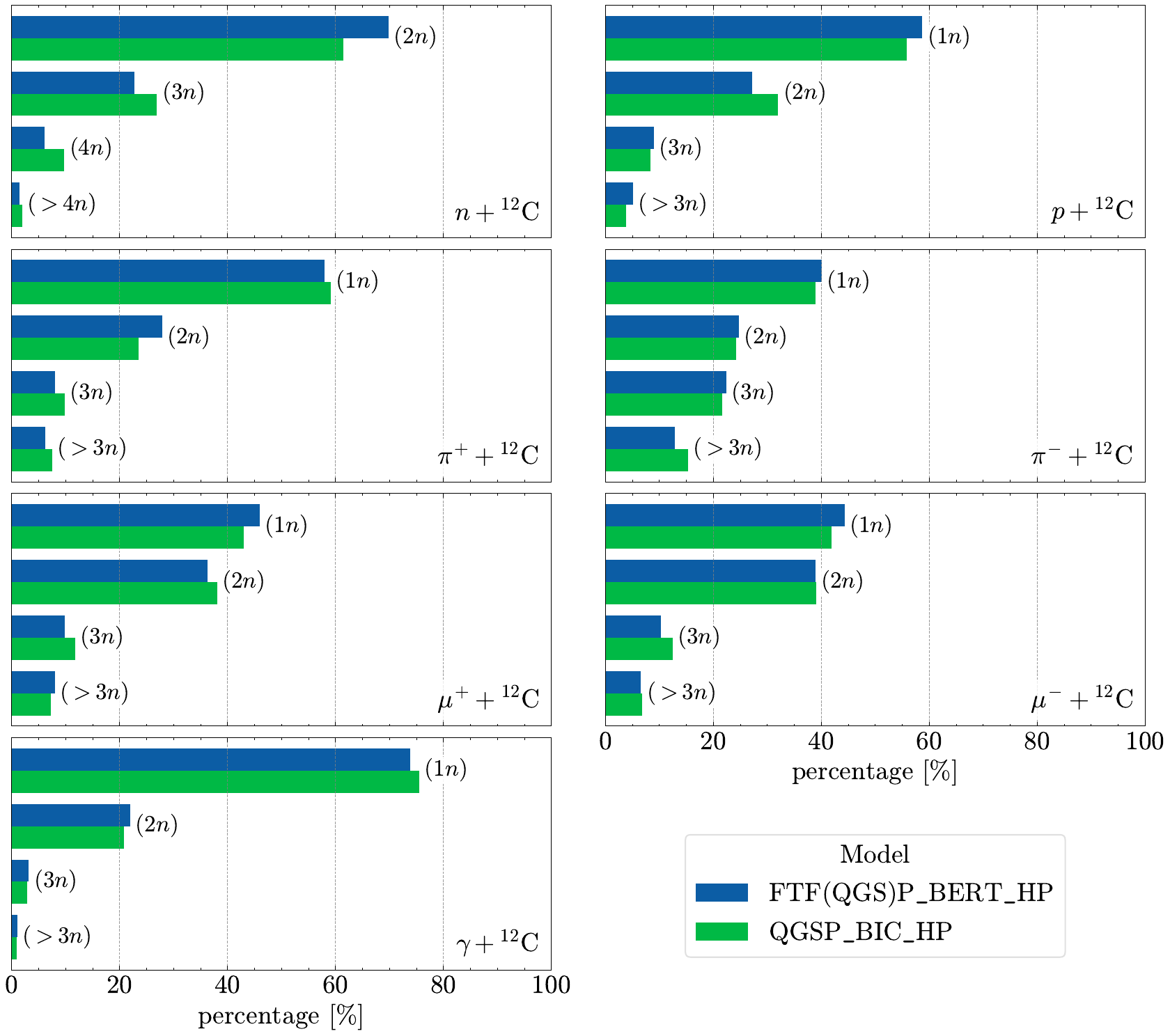}
    \caption{Distributions of exclusive new-neutron-generation reactions on $^{12}\mathrm{C}$ for various parent particles ($n,p,\pi^{+},\pi^{-},\mu^{+},\mu^{-},\gamma$), categorized by the associated final state neutron multiplicities. Events are required to have at least one newly produced neutron. For $n$-induced interactions, event fractions with exactly 2, 3, 4, and $>$4 new neutrons are shown, normalized to 100\%. For all other parent particles categories, fractions are shown for exactly 1, 2, 3, and $>$3 new neutrons, also normalized to $100\%$. Blue histograms represent the nearly identical results from the \textsc{GEANT4} physics lists \texttt{FTFP\_BERT\_HP} and \texttt{QGSP\_BERT\_HP}, while green histograms show the results from \texttt{QGSP\_BIC\_HP} for comparison.}
    \label{fig:particle+12C}
\end{figure}


Complementing the inclusive reaction comparisons for individual parent particles in Figs.~\ref{fig:process_yield_ftfp},\ref{fig:parentKE_allPL}, we investigate how \textsc{GEANT4} hadronic models distribute reaction strength across specific neutron-producing channels. Fig.~\ref{fig:particle+12C} displays the relative final-state neutron multiplicity distributions from new-neutron-generation interactions on a $^{12}\mathrm{C}$ target, categorized by parent-particle types, such as   $n,p,\pi^{+},\pi^{-},\mu^{+},\mu^{-}, \, \rm{and}\, \gamma$. For $n$-induced interactions, event fractions with exactly 2, 3, 4, and $>$4 new neutrons are normalized to 100\%. For all other parent particles, fractions for exactly 1, 2, 3, and $>$3 new neutrons are similarly normalized to 100\%.
Discrepancies in dominant neutron-generation channels between BERT and BIC physics lists for $n-^{12}$C and $p-^{12}$C show that BIC favors final states with higher neutron multiplicity. This is evident in Fig.~\ref{fig:particle+12C}, where differing channel weights across models lead to variations in predicted neutron yields. A direct comparison for $\gamma$, $\pi^{\pm}$ and $\mu^{\pm}$ projectiles is not possible, as the BIC model is not applicable to these interactions on $^{12}$C.


\section{Benchmarking simulations against experimental data}
\label{sec:res}

A key observable for calibrating the hadronic interaction models is the neutron yield, \(Y_{n}\). Following the definition in Ref.~\cite{DayaBay:2017prd}, it is given by
\begin{equation}
    Y_{n} = \frac{N_{n}}{N_{\mu}\,L_{\mathrm{avg}}\,\rho},
    \label{eq:Yn}
\end{equation}
where \(N_{n}\) is the number of neutrons produced in association with \(N_{\mu}\) muons traversing the GdLS target, while \(L_{\mathrm{avg}}\) and \(\rho\) are the average muon path length and the material density, respectively, whose values have been presented in Sects.~\ref{sec:production} and \ref{sec:Calcalation}.
A second key observable is the tagged neutron multiplicity, defined as the number of neutrons captured on Gd following a muon with a nonzero track length in the LS region. This observable is directly comparable with the Daya Bay results~\cite{DayaBay:2017prd} and reflects the complexity of $\mu$-induced spallation cascades. The remainder of this section presents the data--simulation comparison for both observables.

As shown in Sect.~\ref{sec:production}, the physics lists \texttt{FTFP\_BERT\_HP} and \texttt{QGSP\_BERT\_HP} yield statistically identical results for all observables in this work. This agreement is expected, as both employ the same BERT intranuclear cascade model for low- and intermediate-energy hadron transport in the $\mu$-induced spallation regime. To simplify the presentation and focus the comparison on the impact of the intranuclear cascade model, we therefore adopt \texttt{FTFP\_BERT\_HP} as the representative BERT-based list in subsequent figures. Consequently, the data–simulation comparisons in this section are presented only for \texttt{FTFP\_BERT\_HP} and \texttt{QGSP\_BIC\_HP}.

\subsection{Neutron yield}

According to Eq.~\ref{eq:Yn}, we calculate the neutron yield, $Y_{n}^{\rm G}$, for each \textsc{GEANT4} physics list. In this work, the original simulation yield $Y_{n}^{\mathrm{G}}$ is obtained directly from the \textsc{GEANT4} output. The \textsc{TALYS}-based MC yield, $Y_{n}^{\mathrm{T}}$, is evaluated using the same definition but with the raw \textsc{GEANT4} neutron count $N_{n}$ replaced by the corrected count obtained from the event-by-event reweighting scheme detailed in Sect.~\ref{sec:talys}.

The calculated neutron yields are summarized in Tab.~\ref{tab:neutron_yield_compare}. In the original \textsc{GEANT4} simulations, the \texttt{FTFP\_BERT\_HP} physics list yields a neutron yield of about $12.3 \times 10^{-5}\,\mu^{-1}\mathrm{g}^{-1}\mathrm{cm}^{2}$, while \texttt{QGSP\_BIC\_HP} gives a slightly lower value of $11.9 \times 10^{-5}\,\mu^{-1}\mathrm{g}^{-1}\mathrm{cm}^{2}$. Compared to the Daya Bay measurement of $10.26 \times 10^{-5}\,\mu^{-1}\mathrm{g}^{-1}\mathrm{cm}^{2}$~\cite{DayaBay:2017prd}, both original \textsc{GEANT4} lists overestimate the yield by roughly $13\%$–$20\%$.
The application of the \textsc{TALYS}-based reweighting results in a systematic reduction of the neutron yield for both physics lists. As shown in Tab.~\ref{tab:neutron_yield_compare}, the corrected (\textsc{TALYS}-based) yields are 12--14\% lower than the original \textsc{GEANT4} predictions. This adjustment leads to a substantially improved agreement with the experimental measurement. Specifically, for \texttt{FTFP\_BERT\_HP}, the deviation from the Daya Bay result is reduced from approximately $+20\%$ to about $+6\%$. For \texttt{QGSP\_BIC\_HP}, the original overprediction of roughly $+13\%$ is reduced to a level essentially consistent with the data, yielding a residual ratio of \(R_{\text{T/D}}= \dfrac{Y_{n}^{\mathrm{T}} - Y_{n}^{\mathrm{D}}}{Y_{n}^{\mathrm{D}}}\times100\% \approx -0.39\%\).

\begin{table}[htbp]
    \centering
    \setlength{\tabcolsep}{6pt}
    \renewcommand{\arraystretch}{1.15}
    \caption{
    Comparison of simulated neutron yields with the Daya Bay measurement.
    $Y_{n}^{\mathrm{G}}$ denotes the neutron yield from the original \textsc{GEANT4} physics lists, and $Y_{n}^{\mathrm{T}}$ stand for the modified yield based on TALYS hadronic cross sections.
    For reference, the measured neutron yield is $Y_n^{\mathrm{D}} = (10.26 \pm 0.86) \times 10^{-5}\,\mu^{-1}\mathrm{g}^{-1}\mathrm{cm}^{2}$, which is the neutron yield obtained from the Daya Bay experiment in EH1~\cite{DayaBay:2017prd}.
    The relative differences with respect to data are defined as
    $R_{\mathrm{G/D}} = \dfrac{Y_{n}^{\mathrm{G}} - Y_{n}^{\mathrm{D}}}{Y_{n}^{\mathrm{D}}}\times100\%$
    and
    $R_{\mathrm{T/D}} = \dfrac{Y_{n}^{\mathrm{T}} - Y_{n}^{\mathrm{D}}}{Y_{n}^{\mathrm{D}}}\times100\%$.
    }
    \label{tab:neutron_yield_compare}
    \begin{tabular}{lcccc}
        \hline
        \textbf{Physics List} &
        \multicolumn{1}{c}{$Y_{n}^{\mathrm{G}}$} (\(\times10^{-5}\,\mu^{-1}\mathrm{g}^{-1}\mathrm{cm}^{2}\)) &
        \multicolumn{1}{c}{$Y_{n}^{\mathrm{T}}$} (\(\times10^{-5}\,\mu^{-1}\mathrm{g}^{-1}\mathrm{cm}^{2}\)) &
        \multicolumn{1}{c}{$R_{\mathrm{G/D}}$} &
        \multicolumn{1}{c}{$R_{\mathrm{T/D}}$} \\
        \hline
        FTFP\_BERT\_HP & $12.31$ & $10.88$ & $+20.0\%$ & $+6.04\%$ \\
        QGSP\_BIC\_HP  & $11.88$ & $10.22$ & $+13.0\%$ & $-0.39\%$ \\
        \hline
    \end{tabular}
\end{table}

The significant reduction from $Y_{n}^{\mathrm{G}}$ to $Y_{n}^{\mathrm{T}}$ demonstrates the sensitivity of the total neutron yield to inelastic cross sections in the intermediate energy range (50$-$200 MeV). This shift can be understood by examining the differences between \textsc{TALYS} and \textsc{GEANT4} for the dominant new-neutron-generation channels between \textsc{TALYS} and \textsc{GEANT4}, as illustrated in Fig.~\ref{fig:g4_talys_inel_xs_compare}.
While \textsc{TALYS} predicts higher cross sections at specific low energies, its values are systematically lower than \textsc{GEANT4}'s across the broad 50–200 MeV region, where a substantial portion of the secondary cascade flux resides. Consequently, incorporating the more accurate \textsc{TALYS} nuclear reaction probabilities provides a physics-driven correction that effectively counteracts \textsc{GEANT4}'s overprediction. However, it is important to note that the hadronic interaction cross sections provided by \textsc{TALYS} are limited to energies up to 200 MeV and therefore do not extend to the higher-energy regime responsible for the spallation neutron production. 

\subsection{Tagged neutron multiplicity}
\label{sec:neutron-multiplicity}

Following cosmogenic neutron production in the GdLS region, most neutrons undergo capture on Gd, producing an 8 MeV gamma-ray cascade. The Daya Bay analysis~\cite{DayaBay:2017prd} determines the $\mu$-induced neutron yield by selecting neutron captures on Gd that follow a tagged muon signal. In our simulations, we perform three corresponding quantitative evaluations:
\begin{enumerate}
    \item \textbf{Generated neutrons} (\(N_{\mathrm{gen}}\)): number of neutrons produced along $\mu$ tracks within the GdLS.
    \item \textbf{Captured neutrons} (\(N_{\mathrm{nGd}}\)): number of neutrons that capture on Gd, for muons with a nonzero track length in the LS region (see Fig.~\ref{fig:muon_geo}). 
    \item \textbf{Capture fraction} (\(\xi_{\mathrm{cap}}\)): corresponding capture efficiency, defined as $\xi_{\mathrm{cap}} = N_{\mathrm{nGd}} / N_{\mathrm{gen}}$.
\end{enumerate}
Table.~\ref{tab:neutron_cap_compare} summarizes these three key observables for \textsc{GEANT4} \texttt{FTFP\_BERT\_HP}, and \texttt{QGSP\_BIC\_HP} hadronic physics lists.
The capture efficiency exhibits a narrow spread, ranging from 0.67 to 0.69 across these physics lists. This consistency indicates that, conditional on neutron production, the capture probability is governed primarily by factors common to all simulations: the detector geometry, the Gd-doping concentration, and the low-energy neutron transport and capture physics. These common components—specifically, the HP neutron scattering model below $\sim$20 MeV and the neutron-capture physics implemented via \texttt{G4NeutronRadCapture} and \texttt{G4NeutronCaptureXS}—are shared identically across the physics lists~\cite{GEANT4:PLGuide}.

\begin{table}[!tb]
    \centering
    \caption{Comparison of the number of neutrons produced along muon tracks within the GdLS ($N_{\mathrm{gen}}$), the number of neutrons captured on Gd, for muons with a nonzero track length in LS ($N_{\mathrm{nGd}}$), and the corresponding capture efficiency ($\xi_{\mathrm{cap}}$) for the \textsc{GEANT4} hadronic physics lists.}
    \label{tab:neutron_cap_compare}
    \begin{tabular}{lrrr}
    \hline
    \textbf{Physics List} & $N_{\mathrm{gen}}$ & $N_{\mathrm{nGd}}$ & $\xi_{\mathrm{cap}}$ \\
    \hline
    \texttt{FTFP\_BERT\_HP}  & 27,830 & 18,710 & 0.67 \\
    \texttt{QGSP\_BIC\_HP}   & 26,227 & 18,219 & 0.69 \\
    \hline
    \end{tabular}
\end{table}

The tagged neutron multiplicity (\(N_{\mathrm{multi}}\)) for a muon passing through the LS region is extracted directly from the original \textsc{GEANT4} simulation, denoted as \(N_{\mathrm{multi}}^{\mathrm{G}}\). For the \textsc{TALYS}-based MC, it is computed in two steps. 
First, the adjusted number of generated neutrons ($N_{n}^{\rm T}$) for a muon passing through the LS region is calculated using Eq.~\ref{eq:NnT}.
The corresponding tagged neutron multiplicity, \(N_{\mathrm{multi}}^{\mathrm{T}}\), is then obtained by applying the event-specific capture fraction \(\xi_{\mathrm{cap}}\) derived from the \textsc{GEANT4} simulation:
\begin{equation}
    N_{\mathrm{multi}}^{\mathrm{T}} = N_{n}^{\mathrm{T}} \times \xi_{\mathrm{cap}} .
    \label{eq:NcapT}
\end{equation}
This approach relies on the assumption that, for a given muon event, the capture fraction is determined predominantly by the overall event topology and detector geometry, and is largely insensitive to the reweighting of individual vertices. This is an excellent approximation for our setup, as evidenced by the stability of \(\xi_{\mathrm{cap}}\) across physics lists, as shown in Tab.~\ref{tab:neutron_cap_compare}.

\begin{figure}[!tb]
    \centering
    \includegraphics[width=0.48\linewidth]{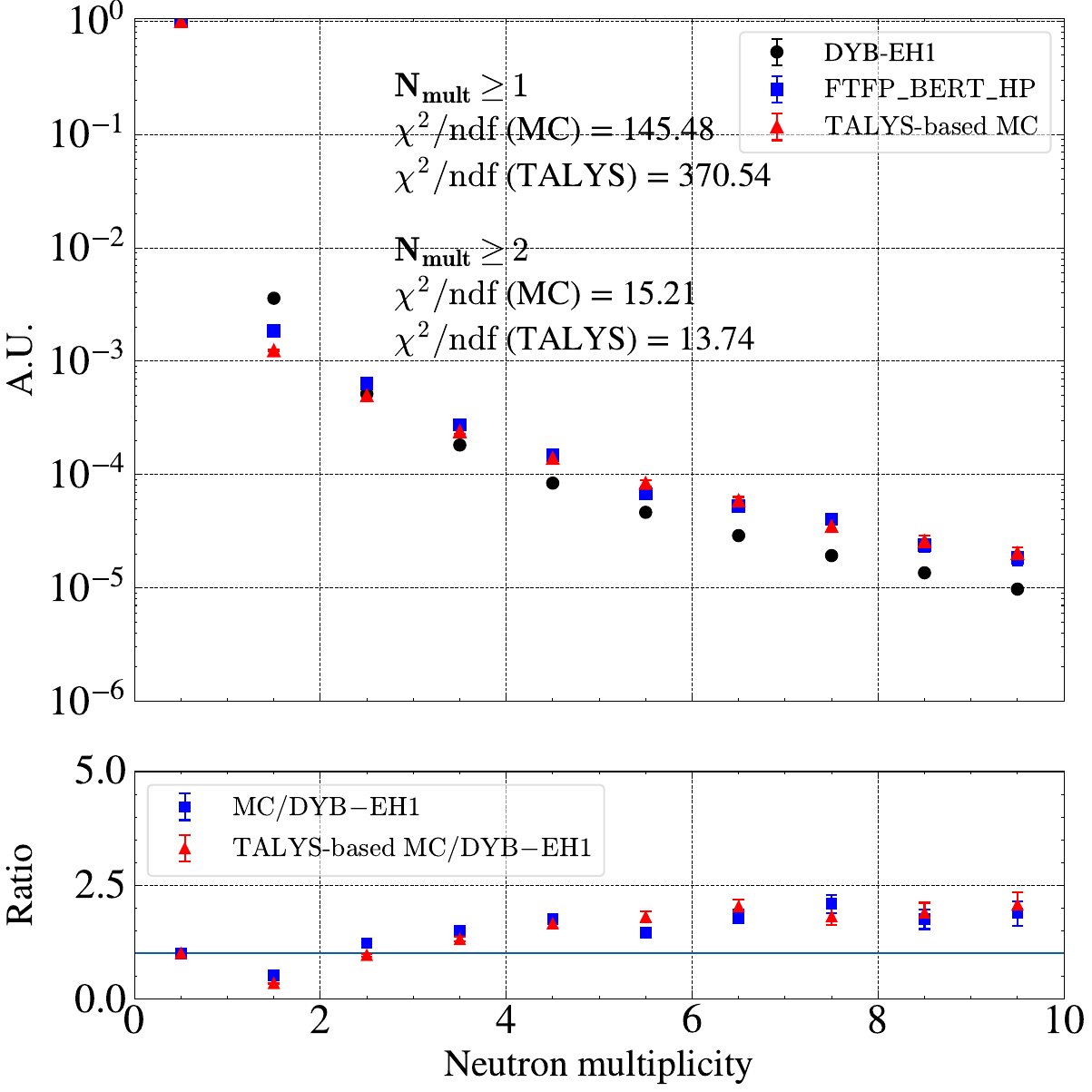}
    \includegraphics[width=0.48\linewidth]{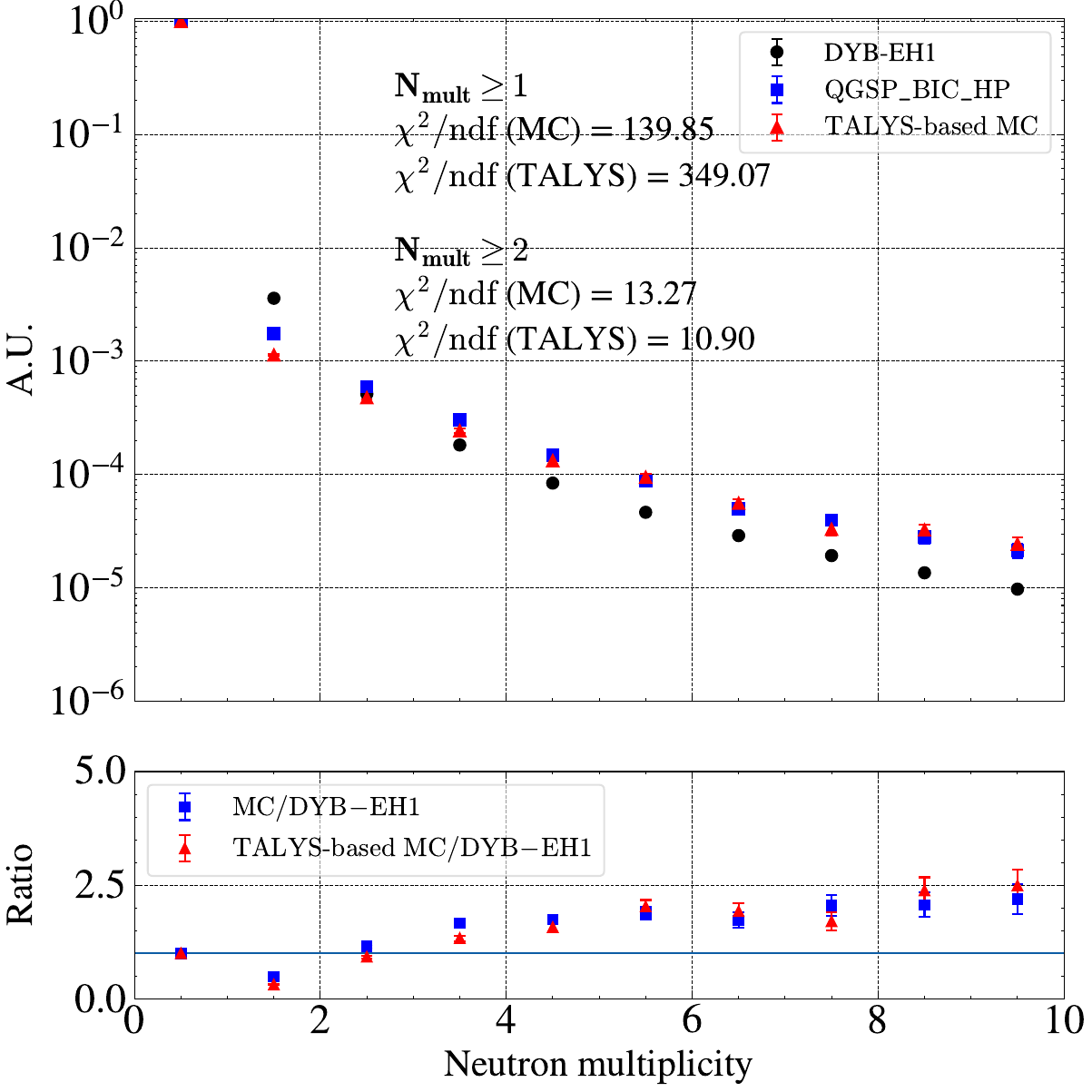}
    \caption{Comparison of the tagged neutron multiplicity distributions between the Daya Bay measurement~\cite{DayaBay:2017prd} and simulations.
    The left (right) panel shows the results for the \texttt{FTFP\_BERT\_HP} (\texttt{QGSP\_BIC\_HP}) physics list.
    In the upper sub-panels, the experimental data from the EH1 LS detector (DYB-EH1) (black markers) are compared with predictions from the original \textsc{GEANT4} simulation (blue markers) and the \textsc{TALYS}-based MC (red markers).
    All spectra are normalized to unit area (A.U.), and the error bars represent statistical uncertainties propagated from Poisson counting.
    Only neutron multiplicities up to $N_{\mathrm{multi}}=10$ are shown, as the contribution from higher-multiplicity events is negligible.
    The lower sub-panels present the bin-by-bin ratios of MC to data.
    The $\chi^{2}/\mathrm{ndf}$ values displayed in the upper panels are computed as $\chi^{2}=\sum_i ({\rm D}_i-{\rm M}_i)^2/(\sigma_{{\rm D},i}^2+\sigma_{{\rm M},i}^2)$ using bins with non-zero contents, and are quoted for $N_{\mathrm{multi}}\geq 1$ and for $N_{\mathrm{multi}}\geq 2$.}
    \label{figure:neutron-mult}
\end{figure}

The top panels of Fig.~\ref{figure:neutron-mult} show the tagged neutron multiplicity distributions from the original \textsc{GEANT4} simulation and the \textsc{TALYS}-based MC, for the \texttt{FTFP\_BERT\_HP} and \texttt{QGSP\_BIC\_HP} physics lists, respectively. To compare with the experimental data, the data from the Daya Bay EH1 LS detector (DYB-EH1)~\cite{DayaBay:2017prd} is also presented in the top panels of Fig.~\ref{figure:neutron-mult}. All spectra are normalized to unit area, and the error bars represent statistical uncertainties only. 
Note that in Ref.~\cite{DayaBay:2017prd}, the comparison of neutron multiplicity between the data and MC in EH1 has been performed. For this comparison, the \textsc{GEANT4} (version 9.2p01) is used. It is found that the difference is largest when neutron multiplicity equals 1, which is the most dominant neutron multiplicity for cosmogenic neutron production. In this work, similar comparisons have been performed in Fig.~\ref{figure:neutron-mult}.   
The bottom panels of Fig.~\ref{figure:neutron-mult} show the ratios of the simulation results to the experimental data. For left and right panels, two ratios are presented: the original \textsc{GEANT4} prediction to data (MC/DYB-EH1) and the \textsc{TALYS}-based adjusted prediction to data (\textsc{TALYS}-based MC/DYB-EH1). To further assess the consistency between data and simulations with different hadronic models, we quantify the goodness-of-fit using a bin-by-bin $\chi^2$ test that incorporates statistical uncertainties. 
The \(\chi^{2}\) is calculated using bins with non-zero content according to
\begin{equation}
    \chi^{2} = \sum_{i} \frac{({\rm D}_i - {\rm M}_i)^2}{\sigma_{{\rm D},i}^{2} + \sigma_{{\rm M},i}^{2}},
    \label{eq:chi2_def}
\end{equation}
where \({\rm D}_i\) and \({\rm M}_i\) denote the content of the \(i\)-th bin for the data and the simulation, respectively, and \(\sigma_{{\rm D},i}\) and \(\sigma_{{\rm M},i}\) are their corresponding statistical uncertainties.
The goodness of fit is quantified by the \(\chi^{2}\) per degree of freedom (\(\chi^{2}/\mathrm{ndf}\)). Values are reported separately for two event selections: those with neutron multiplicity \(N_{\mathrm{multi}} \geq 1\) and those with \(N_{\mathrm{multi}} \geq 2\). For each selection, we quote \(\chi^{2}/\mathrm{ndf}\) for the original \textsc{GEANT4} simulation (denoted \(\chi^{2}/\mathrm{ndf}\)(MC)) and for the \textsc{TALYS}-based MC prediction (denoted \(\chi^{2}/\mathrm{ndf}\)(\textsc{TALYS})).
Key observations from Fig.~\ref{figure:neutron-mult} are summarized as follows.

\begin{itemize}
    \item \textbf{Neutron multiplicity $N_{\mathrm{multi}} = 1$}: All simulation predictions underestimate the experimental data. Applying the \textsc{TALYS}-based correction further enlarges this discrepancy for both physics lists. This is reflected in the $\chi^{2}/\mathrm{ndf}$ values: for events with $N_{\mathrm{multi}} \geq 1$, the $\chi^{2}/\mathrm{ndf}$ from the \textsc{TALYS}-based MC is worse than that from the original \textsc{GEANT4} prediction. 

    \item \textbf{Neutron multiplicities $N_{\mathrm{multi}} > 1$}: While all simulations systematically overestimate the data in this region, the \textsc{TALYS}-based MC spectra show slightly better agreement. The improvement is most pronounced in the multiplicity range ($N_{\mathrm{multi}} = 2$--$8$), where the original \textsc{GEANT4} predictions exhibit a marked tendency to overproduce multi-neutron events.

    \item \textbf{Comparison of hadronic models}: For all multiplicities, the original \textsc{GEANT4} prediction using the BIC-based model agrees better with data than the prediction using the BERT-based model. The corresponding \textsc{TALYS}-based MC corrections preserve this relative behavior. This conclusion is also supported by the comparison of the neutron yield.
\end{itemize}

\begin{figure}[!tb]
    \centering
    \includegraphics[width=0.8\linewidth]{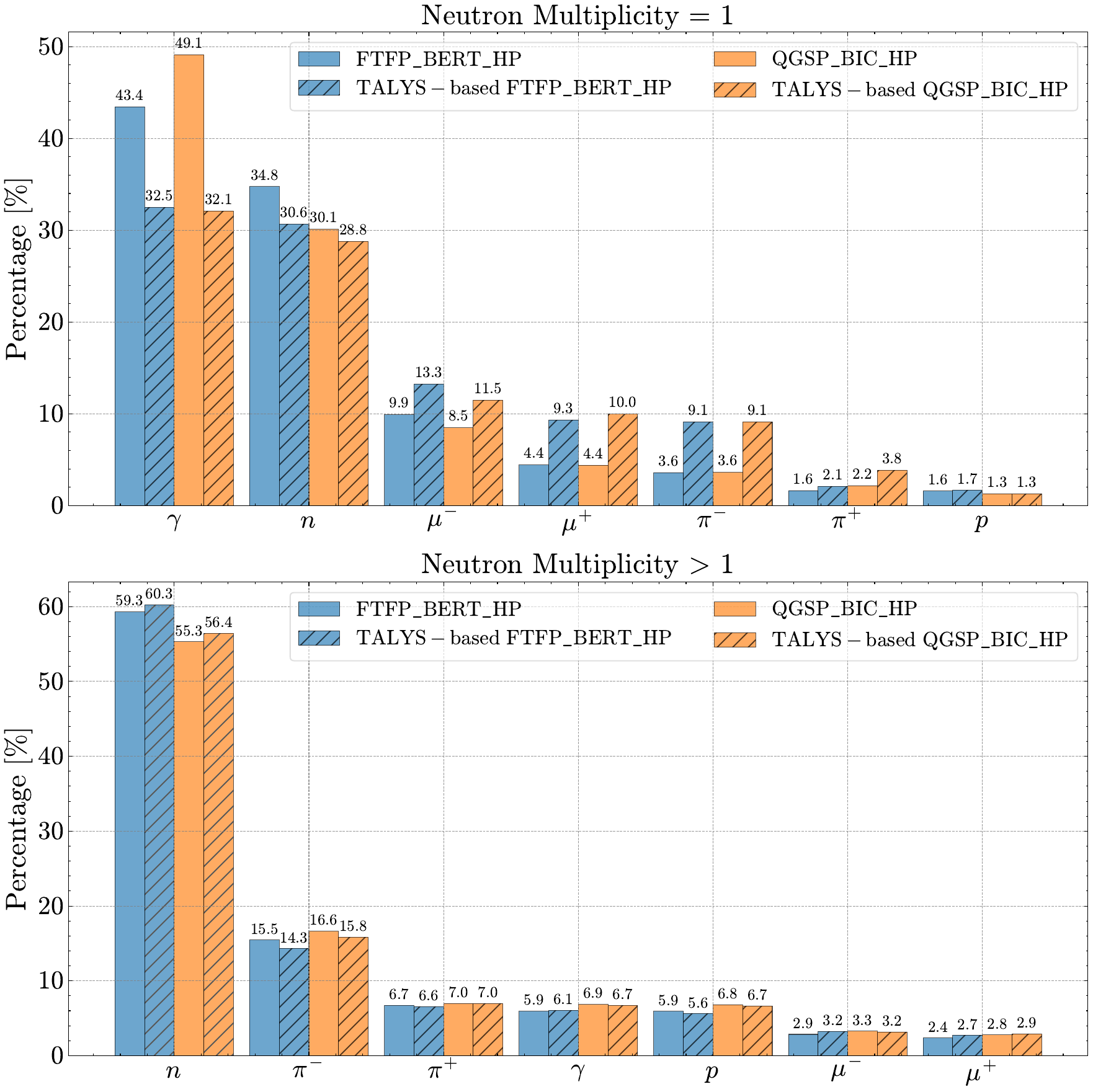}
    \caption{
        Parent particle contributions to neutron-producing events for the original \textsc{GEANT4} physics lists
        \texttt{FTFP\_BERT\_HP}, \texttt{QGSP\_BIC\_HP} and their \textsc{TALYS}-based MC variants.
        The upper panel shows events with exactly one tagged neutron ($N_\mathrm{multi} = 1$),
        while the lower panel corresponds to events with multiple tagged neutrons ($N_\mathrm{multi} > 1$).
        For each parent species ($\gamma$, $n$, $\mu^\pm$, $\pi^\pm$, $p$), the relative fractions are compared side by side, illustrating the redistribution of contributions in the \textsc{TALYS}-based MC.
    }
    \label{figure:neutron-mult-splite}
\end{figure}

To investigate the composition of events with $N_{\mathrm{multi}} = 1$ and $N_{\mathrm{multi}} > 1$ in detail, 
Fig.~\ref{figure:neutron-mult-splite} presents the relative contributions of various parent particles to these event classes. We compare the original \textsc{GEANT4} physics lists (\texttt{FTFP\_BERT\_HP} and \texttt{QGSP\_BIC\_HP}) with their \textsc{TALYS}-based counterparts.
The results indicate that single-tagged-neutron events are predominantly induced by $\gamma-^{12}$C and $n-^{12}$C reactions, whereas multi-tagged-neutron events primarily originate from secondary neutron interactions. A comparison between the physics lists shows that \texttt{QGSP\_BIC\_HP} predicts a higher fraction of $\gamma$-induced single-tagged-neutron events and a slightly lower fraction of $n$-induced multi-tagged-neutron events relative to \texttt{FTFP\_BERT\_HP}. 

Given the superior agreement of the original BIC-based \textsc{GEANT4} model with the data across all multiplicities compared to the BERT-based model, a promising strategy for reducing the remaining data--MC discrepancy is to adjust the key inelastic cross sections. Specifically, this would involve increasing the $\gamma$--\(^{12}\)C cross section for single-tagged-neutron events. 
This direction is partially validated by the application of \textsc{TALYS}-based hadronic cross sections. In the \textsc{TALYS}-based MC, the fraction of $\gamma$--\(^{12}\)C reactions contributing to events with \(N_{\mathrm{multi}} = 1\) is lower than in the original \textsc{GEANT4} simulation, which worsens the agreement with data for this channel.

\section{Summary}

This work presents a benchmark study of cosmogenic neutron production in large LS detectors. We perform detailed simulations based on the Daya Bay experiment using three \textsc{GEANT4} hadronic physics lists—\texttt{FTFP\_BERT\_HP}, \texttt{QGSP\_BERT\_HP}, and \texttt{QGSP\_BIC\_HP}—and develop an innovative methodology for adjusting interaction cross sections based on \textsc{TALYS} nuclear data. The discrepancies between the Daya Bay measurements and the simulations are quantified by comparing both the neutron yield and the tagged neutron multiplicity distributions. The key findings are summarized as follows.
\begin{itemize}
    \item  Differences exist among the various \textsc{GEANT4} physics lists in their predictions of muon-induced spallation neutron yields, underscoring the significant model dependency in related background simulations. For cosmic muons with an average energy of \(\sim 64\ \mathrm{GeV}\), where secondary particles predominantly reside in the sub-GeV kinetic energy regime, the final neutron yield is primarily governed by the choice of the low-energy intranuclear cascade model---either BERT or BIC---rather than by the high-energy string models (FTF or QGS). This work provides a detailed investigation of the impact of the BERT and BIC models on cosmogenic neutron production. A comparison with experimental data reveals that the BIC model yields predictions more consistent with the measurements than the BERT model.

    \item Rescaling the exclusive inelastic cross sections for neutrons, protons, and gamma rays using \textsc{TALYS}--\textsc{GEANT4} ratios systematically improves the agreement of the total neutron yield, \(Y_n\), with the experimental data. 
    For the BERT-based models, the discrepancy with the Daya Bay measurement is reduced from approximately 20\% to about 6\%, while for the BIC-based models it is further reduced from roughly 13\% to the sub-percent level ($\sim$0.3\%).
    However, it should be noted that the hadronic interaction cross sections provided by \textsc{TALYS} are limited to energies up to 200~MeV and therefore do not extend to the higher-energy regime relevant for the full spallation neutron yield. For future studies, the inclusion of additional nuclear data at higher energies will be essential to further constrain the hadronic interaction models.

    \item An analysis of the neutron multiplicity distributions reveals that the most significant discrepancy between the experimental data and our simulations persists for single-tagged-neutron events (\(N_{\mathrm{multi}} = 1\)). This deficit was already identified in a previous comparison that employed \textsc{GEANT4} (version 9.2p01). A quantitative \(\chi^{2}\) test further demonstrates that predictions from the BIC-based models show better overall agreement with the data. 
    
    \item Compared to the improvement achieved in the neutron yield with the \textsc{TALYS}-based correction, the adjustment has a differential impact on the tagged neutron multiplicity distributions. While the agreement for multi-neutron events (\(N_{\mathrm{multi}} > 1\)) improves, a persistent deficit in the yield of single-neutron events remains. Consequently, when single-neutron events are included, the overall \(\chi^{2}\)/ndf deteriorates. This indicates that while the \textsc{TALYS}-based correction successfully addresses normalization issues stemming from exclusive cross sections, it unveils a residual discrepancy in the final-state modeling, particularly for channels leading to a single final-state neutron.

    \item For events with \(N_{\mathrm{multi}} = 1\), neutron production is predominantly due to the \(\gamma\)--\(^{12}\mathrm{C}\) reaction. The relative contribution of this channel is higher in the original BIC-based models than in the BERT-based models or their \textsc{TALYS}-corrected counterparts. Given the superior agreement of the original BIC-based \textsc{GEANT4} model with the data, a promising strategy for reducing the remaining data--MC discrepancy is to increase the inelastic cross section for the \(\gamma\)--\(^{12}\mathrm{C}\) channel. Conversely, the persistent deficit in simulated single-neutron events indicates that current implementations of the Fermi break-up and evaporation models in \textsc{GEANT4} require further refinement to accurately reproduce the experimental multiplicity distributions.
\end{itemize}

Building on these key findings, we propose a practical two-step strategy for improving background modeling in large LS detectors (e.g., JUNO). First, apply \textsc{TALYS}-based cross section adjustments to obtain an accurate normalization of the total cosmogenic neutron yield for background subtraction. Second, use the residual discrepancies in the neutron multiplicity distributions---particularly the deficit in the \(N_{\mathrm{multi}} = 1\) channel---as constraints for the future refinement of intranuclear cascade and deexcitation models. Note that the proposed strategy is applicable not only to LS detectors but also to other large-volume neutrino detectors---such as water Cherenkov detectors (e.g., Hyper-Kamiokande~\cite{Hyper-Kamiokande:2018ofw}) and liquid-argon time-projection chambers (e.g., DUNE~\cite{DUNE:2020jqi})---where accurate modeling of cosmogenic neutron backgrounds is equally critical.

This study establishes a reproducible framework for modeling spallation neutrons and provides a clear strategy for mitigating persistent data--simulation discrepancies. This is achieved through channel-specific cross-section reweighting and targeted parameter scans of intranuclear cascade models. The developed methodology offers a template for the data-driven refinement of hadronic interaction models. Its potential applications extend beyond cosmogenic neutrons to analogous background challenges, such as those posed by atmospheric neutrino neutral-current (NC) interactions~\cite{Cheng:2020aaw,Cheng:2020oko,Cheng:2024uyj,Super-Kamiokande:2025cht,Super-Kamiokande:2023oxd}. Notably, the energy range for cosmogenic neutron production examined in this work encompasses that of the final-state particles generated in such NC interactions.

\section*{Acknowledgements}

The authors would like to thank Yufeng Li for carefully reading the manuscript and useful comments. 
This work is supported in part by National Natural Science Foundation of China under Grant Nos. ~12405125, ~12575210, by the National Key R\&D Program of China under Grant No. ~2024YFE0110500.

\label{sec:sum}

\end{document}